\documentclass[prl,twocolumn,showpacs,superscriptaddress,preprintnumbers]{revtex4}
\usepackage{graphicx}
\usepackage{dcolumn}
\usepackage{amsmath}
\usepackage{color}

\begin{document}

\title{
Electronic correlations at the $\alpha$--$\gamma$ structural phase transition in paramagnetic iron}

\author{I. Leonov}
\affiliation{Theoretical Physics III, Center for Electronic Correlations and Magnetism,
Institute of Physics, University of Augsburg, Augsburg 86135, Germany}
\author{A. I. Poteryaev}
\affiliation{Institute of Metal Physics, S. Kovalevskoy St. 18, 620219 Yekaterinburg
GSP-170, Russia}
\affiliation{Institute of Quantum Materials Science, 620107 Yekaterinburg,
Russia}
\author{V. I. Anisimov}
\affiliation{Institute of Metal Physics, S. Kovalevskoy St. 18, 620219 Yekaterinburg
GSP-170, Russia}
\author{D. Vollhardt}
\affiliation{Theoretical Physics III, Center for Electronic Correlations and Magnetism,
Institute of Physics, University of Augsburg, Augsburg 86135, Germany}

\begin{abstract}
We compute the equilibrium crystal structure and phase stability of iron at the $\alpha$(bcc)--$\gamma$(fcc)
phase transition as a function of temperature, by employing a combination of \textit{ab initio} methods for
calculating electronic band structures and dynamical mean-field theory. The magnetic correlation energy is
found to be an essential driving force behind the $\alpha$--$\gamma$ structural phase transition in
paramagnetic iron.
\end{abstract}

\pacs{71.10.-w, 71.27.+a} \maketitle

The properties of iron have fascinated mankind for several thousand years already. Indeed, iron has been
an exceptionally important material for the development of modern civilization and its technologies.
Nevertheless, even today many properties of iron, e.g., at high pressures and temperatures, are still
not sufficiently understood. Therefore iron remains at the focus of active research.

\begin{figure}[tbp!]
\centerline{\includegraphics[width=0.45\textwidth,clip=true]{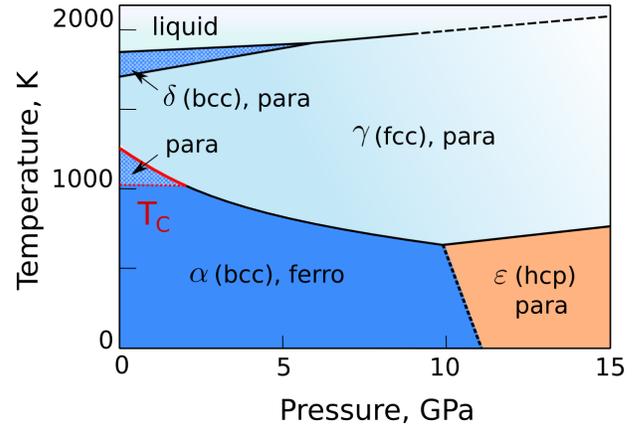}}
\caption{(color online)
Schematic temperature-pressure phase diagram of iron (see text).
} \label{fig:dgrm}
\end{figure}

At low pressures and temperatures iron crystallizes in a body-centered cubic (bcc) structure,
referred to as $\alpha$-iron or ferrite; see Fig.~\ref{fig:dgrm}. In particular, at ambient pressure iron is ferromagnetic, with
an anomalously high  Curie temperature of $T_C \sim 1043$ K. Upon heating, iron exhibits several structural
phase transformations \cite{BH55,W63}: at $\sim$ 1185 K to the face-centered cubic (fcc) phase ($\gamma$-iron
or austenite), and at $\sim$ 1670 K again to a bcc structure ($\delta$-iron). At high
pressure iron becomes paramagnetic with a hexagonal close packed structure ($\epsilon$-iron).

Density functional theory (DFT) in the
local spin density approximation gives a quantitatively accurate description of the ordered magnetic
moment and the spin stiffness of bcc-Fe \cite{S98}, but predicts the nonmagnetic fcc structure to be more
stable than the observed ferromagnetic bcc phase \cite{WKK85}. Only if the spin polarized
generalized-gradient approximation (GGA) \cite{PB96} is applied does one obtain the correct ground
state properties of iron \cite{SP91}.
%

Stoner theory of ferromagnetism \cite{S39} can give a qualitatively correct description of several
magnetic and structural properties of iron, but predicts a simultaneous magnetic and structural change
at the bcc--fcc phase transition with a local moment collapse while, in fact, the bcc--fcc phase
transition occurs $\sim 200$ K \emph{above} $T_C$; see Fig.~\ref{fig:dgrm}.
Clearly, to account for finite temperature effects of itinerant magnets one requires a formalism which
takes into account the existence of local moments above $T_C$. While the spin-fluctuation theory, which
describes the paramagnetic state above $T_C$ as a collection of disordered moments, 
gives an overall good qualitative explanation of the pressure-temperature phase diagram of iron \cite{HP83}
it fails to provide a reasonably quantitative description and, in particular, predicts the bcc-fcc phase
transition to occur below $T_C$.


The LDA+DMFT computational scheme \cite{LDA+DMFT}, a combination of the DFT in the local density
approximation (LDA) with dynamical mean-field theory (DMFT) \cite{DMFT}, goes beyond the approaches
discussed above since it explicitly includes many-body effects in a non-perturbative and thermodynamically
consistent way. LDA+DMFT was already used to calculate the magnetization and the susceptibility of $\alpha$-iron
as a function of the reduced temperature $T/T_C$ \cite{LK01}.
The calculations gave overall good agreement with experimental data.
The problem has been recently revisited by Katanin \emph{et al.} \cite{KPE10}
who found that the formation of local moments in paramagnetic $\alpha$-Fe
is governed by the $e_g$ electrons and is accompanied by non-Fermi liquid behavior.
This supports the results obtained with the $s$--$d$ model for the $\alpha$-phase of iron \cite{IKT93}.

A recent implementation of the LDA/GGA+DMFT scheme in plane-wave pseudopotentials \cite{LB08,TL08}
now allows one to investigate correlation induced lattice transformations such as the
cooperative Jahn-Teller distortion in KCuF$_3$ and LaMnO$_3$.
The method was not yet used to study structural phase transitions in a paramagnetic
correlated electron system with temperature (or pressure) involving a change of symmetry.
This will be the goal of the present investigation.

\begin{figure}[tbp!]
\centerline{\includegraphics[width=0.45\textwidth,clip]{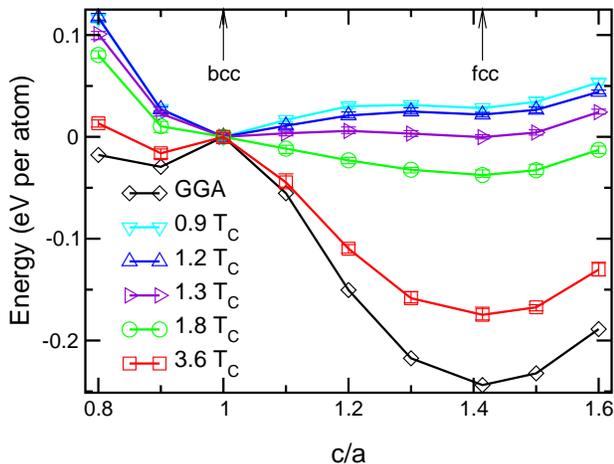}}
\caption{(color online)
Variation of the total energy of paramagnetic iron obtained by GGA and
GGA+DMFT(QMC) for different temperatures. The total energy is calculated
along the bcc-fcc Bain transformation path with constant volume
($a=2.91$ \AA\ for the bcc phase). Error bars indicate
the statistical error of the DMFT(QMC) calculations. }
\label{fig:energy}
\end{figure}

In this Letter we employ the above-mentioned implementation of the LDA/GGA+DMFT scheme \cite{LB08,TL08}
to explore the structural and magnetic properties of paramagnetic iron at finite temperatures. In particular, we will
study the origin of the $\alpha$--$\gamma$ structural phase transformation, and the importance of electronic
correlations for this transition.
We first compute the nonmagnetic GGA electronic structure of iron 
\cite{PSEUDO}. 
To model the bcc-fcc phase transition we employ the Bain transformation
path which is described by a single structural parameter $c/a$, the uniaxial deformation along [001] axis,
with $c/a=1$ for the bcc and $c/a=\sqrt 2$ for the fcc structure.
Here the lattice volume is kept at the experimental volume of $\alpha$-iron
($a=2.91$ \AA) \cite{W63}
in the vicinity of the bcc-fcc phase transition,
while the $c/a$ ratio is changed from 0.8 to 1.6.
Overall, the GGA results qualitatively agree with previous band-structure calculations \cite{SP91}.
In particular, the nonmagnetic GGA yields the fcc structure to be more energetically favorable than the
bcc one (see Fig.~\ref{fig:energy}).


Next we apply the GGA+DMFT approach \cite{LB08,TL08} to determine the structural phase stability of iron.
For the partially filled Fe $sd$ orbitals we construct a basis of atomic-centered symmetry-constrained
Wannier functions \cite{TL08}. The corresponding first-principles multiband Hubbard Hamiltonian has the
form
\begin{eqnarray}
{\hat H} & = & {\hat H_\mathrm{GGA}} + \frac{1}{2}\sum_{imm',\sigma\sigma'}
U^{\sigma \sigma'}_{mm'} \hat{n}_{im\sigma} \hat{n}_{im'\sigma'} - {\hat H_\mathrm{DC}}
\label{eqn:hamilt}
\end{eqnarray}
where $\hat n_{im\sigma} = \hat c^\dagger_{im\sigma} \hat c_{im\sigma}$
and $\hat c^\dagger_{im\sigma}$ ($\hat c_{im\sigma}$) creates (destroys)
an electron with spin $\sigma$ in the Wannier orbital $m$ at site $i$.
Here ${\hat H_\mathrm{GGA}}$ is the effective low-energy Hamiltonian in the basis of Fe $sd$ Wannier orbitals.
The second term on the right-hand side
of Eq.~\ref{eqn:hamilt} describes the Coulomb
interaction between Fe $3d$ electrons in the density-density approximation. It is expressed in terms
of the average Coulomb repulsion $U$ and Hund's rule exchange $J$. In this calculation we use $U=1.8$ eV
which is within the theoretical and experimental estimations $\sim 1-2$ eV and $J=0.9$ eV \cite{U_Fe}.
Further,
${\hat H_\mathrm{DC}}$ is a double counting correction which accounts
for the electronic interactions already
described by the GGA (see below).


In order to identify correlation induced structural transformations, we calculate \cite{LB08} the
total energy as
\begin{eqnarray}
\label{eq:etot}
E &=& E_\mathrm{GGA}[\rho] + \langle {\hat H_\mathrm{GGA}} \rangle - \sum_{m,k}
\epsilon^\mathrm{GGA}_{m,k}  \notag
\\ &+& \frac{1}{2} \sum_{imm',\sigma\sigma'}
U^{\sigma \sigma'}_{mm'} \langle \hat n_{im\sigma} \hat n_{im'\sigma'}\rangle - E_\mathrm{DC},
\end{eqnarray}
where $E_\mathrm{GGA}[\rho]$ denotes the total energy obtained by GGA. Here
$\langle {\hat H_\mathrm{GGA}} \rangle$ is evaluated as the thermal average of the GGA
Wannier Hamiltonian. The third term on the right-hand side
of Eq.~(\ref{eq:etot}) is the
sum of the Fe $sd$ valence-state eigenvalues.
The interaction energy, the 4-th term on the right-hand side of Eq.~(2),
$\langle \hat n_{im\sigma} \hat n_{im'\sigma'}\rangle$ which is calculated in DMFT.
The double-counting correction $E_\mathrm{DC}= \frac{1}{2} \sum_{imm',\sigma\sigma'}
U^{\sigma \sigma'}_{mm'} \langle \hat n_{im\sigma} \rangle \langle \hat n_{im'\sigma'} \rangle$
corresponds to the average Coulomb repulsion between electrons in the Fe $3d$ Wannier orbitals
calculated from the self-consistently determined local occupancies \cite{dcc}.


To solve the realistic many-body Hamiltonian (1) within DMFT we employ quantum Monte Carlo (QMC)
simulations with the Hirsch-Fye algorithm \cite{HF86}. The calculations for iron are performed
along the Bain transformation path as a function of the reduced temperature $T/T_C$.
Here $T_C$  corresponds to the temperature where the spin polarization in the self-consistent
GGA+DMFT solution vanishes. We obtain $T_C\sim1600$ K which, given the local nature of the DMFT
approach, is in reasonable agreement with the experimental value of 1043 K and also with earlier
LDA+DMFT calculations \cite{LK01}. We find that $T_C$ depends sensitively on the lattice distortion
$c/a$. It has a maximum value for the bcc ($c/a=1$) structure and decreases rapidly for other values.
In particular, for all temperatures considered here the fcc phase remains paramagnetic.


In Fig.~\ref{fig:energy} we show the variation of the total energy of paramagnetic iron with
temperature along the bcc-fcc Bain transformation path.
The result exhibits two well-defined energy minima at $c/a = 1$ (at low temperature) and
$c/a=\sqrt 2$ (at high temperature), corresponding to the bcc and fcc structures, respectively.
We find that for decreasing temperature the inclusion of the electronic correlations among the
partially filled Fe $3d$ states considerably reduces the total energy difference between
the $\alpha$ and $\gamma$ phases.
In particular, the bcc-to-fcc structural phase transition is found to take place
at $T_{\rm struct}\sim 1.3~T_C$, i.e., well \emph{above} $T_C$ \cite{entropy_Fe}. Our result for
$\Delta T \equiv T_{\rm struct}-T_C$, the difference between the temperatures at which the magnetic
transition and the structural phase transition occur, is in remarkable agreement with
the experimental result of $\Delta T \sim 200$ K. This finding differs
from conventional band-structure calculations which predict the magnetic and structural phase
transition to occur simultaneously.
Both $T_{\rm struct}$  and $T_C$ vary sensitively with the value of the Coulomb repulsion $U$
employed in GGA+DMFT calculation. We find that $T_{\rm struct}$ increases for increasing $U$
values, whereas $T_C$ decreases, in agreement with the Kugel-Khomskii theory \cite{KH82}.

In addition, we performed LDA+DMFT calculations to determine the phase stability of iron
at the bcc-fcc phase transition as a function of temperature. 
In contrast to the standard band structure approach where it is
essential that the spin-polarized GGA is used to obtain the correct ground state properties
of iron, we find that both the LDA+DMFT and GGA+DMFT schemes give qualitatively similar results.
In particular, both schemes find the bcc-to-fcc structural phase transition at $\sim 1.3~T_C$,
i.e., well above the magnetic transition. Explanations of the bcc-fcc structural phase transition
and the fact that $T_{\rm struct} \neq T_C$ obviously need to go beyond conventional band structure theories.
This clearly demonstrates the crucial importance of the electronic correlations among the
partially filled Fe $3d$ states.


\begin{table}[tbp!]
\caption{ Calculated lattice constant $a$ for the bcc lattice,
volume $V$ and bulk modulus $B$ for the equilibrium phase of
paramagnetic iron as a function of $T/T_C$.
}
\begin{ruledtabular}
\begin{tabular}{lcccc}
$T/T_C$ & Eq. phase & $a$, \AA\ & $V$, au$^3$ & $B$, Mbar \\
\hline
0 {\scriptsize (GGA)}  & bcc  & 2.757 & 70.71  &  2.66 \\
                       & fcc  & 2.737 & 69.20  &  2.82 \\
0.9                    & bcc  & 2.880 & 80.64  &  1.48 \\
1.2                    & bcc  & 2.883 & 80.84  &  1.50 \\
1.4                    & fcc  & 2.861 & 79.03  &  1.61 \\
1.8                    & fcc  & 2.862 & 79.13  &  1.59 \\
Exp\cite{BH55,W63,BulkModulus}     & bcc/fcc  & 2.88-2.91  &   &  1.62-1.76   \\
\end{tabular}
\end{ruledtabular}
\label{tab:tab1}
\end{table}

Next we perform a structural optimization and compute the equilibrium volume and the
corresponding bulk modulus of paramagnetic iron (see Table~\ref{tab:tab1}).
The bulk modulus is calculated as the derivative of the total energy as a function of volume.
We find that at the bcc-fcc phase transition
the equilibrium lattice volume simultaneously shrinks by $\sim 2$ \%, a result which
is in good agreement with the experimental value of $\sim 1$ \% \cite{BH55}.
The volume reduction is accompanied by an increase of the calculated bulk modulus.
Overall, the equilibrium volume and bulk modulus computed by GGA+DMFT agree well
with the experimental data \cite{BH55,W63,BulkModulus}.

\begin{figure}[tbp!]
\centerline{\includegraphics[width=0.45\textwidth,clip=true]{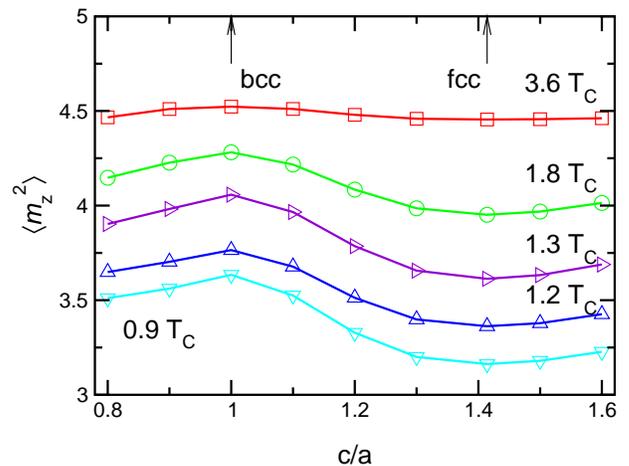}}
\caption{(color online)
Variation of the square of the local magnetic moment
calculated by GGA+DMFT for paramagnetic iron. } \label{fig:muz}
\end{figure}

Finally we compute the square of the instantaneous local moment
$\langle m_z^2 \rangle=\langle(\sum_m [\hat n_{m\uparrow} - \hat n_{m\downarrow}])^2 \rangle$
of paramagnetic iron for the distortions $c/a$ considered here. In Fig.~\ref{fig:muz} we show
the result plotted for various temperatures. At low temperatures, the squared local moment
depends quite strongly  on the value of $c/a$, and is maximal in the bcc and minimal in the
fcc phase, respectively. As expected, above $T_C$, the square local moment gradually increases
with temperature and becomes essentially independent of $c/a$, as indicated by the curve for 
$T=3.6~T_C$ in Fig.~\ref{fig:muz} (we note that this is only a hypothetical curve since at such 
an elevated temperatures iron is already in its liquid state).
This finding has important implications for our understanding of the actual driving force
behind the bcc-to-fcc paramagnetic phase transition. For this we note that the squared
local moment $\langle m_z^2 \rangle$ determines the magnetic correlation energy
$-\frac{1}{4}I \langle m_z^2 \rangle$, which is an essential part \cite{eqHUI}
of the total correlation energy of the Hamiltonian (1).
At high temperatures, when the local moment is almost independent of $c/a$ and the GGA+DMFT
approach finds the fcc phase to be stable, the contribution of the magnetic correlation energy
to the bcc-fcc total energy difference is seen to be negligible. This changes markedly when the
temperature is lowered. Namely, upon cooling the contribution of the magnetic correlation energy
gradually increases and becomes strong enough to overcome the DMFT kinetic energy loss
$E_{kin}= E_\mathrm{GGA}[\rho] + \langle {\hat H_{GGA}} \rangle - \sum_{m,k} \epsilon^\mathrm{GGA}_{m,k}$
for the bcc phase as compared with the fcc phase. Thereby the bcc phase with its larger value of
the local moment is stabilized at $T < 1.3~T_C$. We therefore conclude that the bcc-to-fcc
paramagnetic phase transition is driven by the magnetic correlation energy.

In conclusion, we employed the GGA+DMFT many-body approach to compute the equilibrium
crystal structure and phase stability of iron at the bcc-fcc transition. In particular,
we found that the bcc-to-fcc structural phase transition occurs well above the magnetic
transition, and that the \emph{magnetic} correlation energy is essential to explain this
\emph{structural} transition in paramagnetic iron. The above result and those for the
equilibrium lattice constant and the variation of the unit cell volume at the bcc-fcc phase
transition agree well with experiment.

\begin{acknowledgments}
We thank J. Deisenhofer, Yu. N. Gornostyrev, F. Lechermann, A. I. Lichtenstein,
M. I. Katsnelson, A. A. Katanin, K. Samwer,
and V. Tsurkan for valuable discussions. Support by the Russian Foundation for Basic
Research under Grant No. RFFI-07-02-00041, RFFI-08-02-91953, the Deutsche Forschungsgemeinschaft
through SFB 484 in 2009 and TRR 80 as of January 1, 2010, is gratefully acknowledged.

\end{acknowledgments}


\begin{thebibliography}{99}
\bibitem{dcc}
We note that using a double counting correction in the fully localized limit gives
qualitatively the same results for the phase stability of paramagnetic Fe.
\bibitem{eqHUI}
Neglecting the
slight difference between the Fe $t_{2g}$ and $e_g$ occupancies,
the Coulomb interaction part of the Hamiltonian (1) can be rewritten as
$\hat{H}_{U} = \frac{1}{2}\bar{U} \hat{N}^2 - \frac{1}{4}I \hat{m}_z^2$, with an averaged
Coulomb interaction $\bar{U} \sim 0.75$ eV and exchange interaction
$I=\frac{1}{5}(U+4J) \sim$ 1.08 eV.
%
Here $\hat{N}=\sum_{m\sigma}\hat{n}_{im\sigma}$ and
$\hat{m}_z=\sum_m (\hat{n}_{im\uparrow} - \hat{n}_{im\downarrow})$ are operators of the
total number of particles and magnetization, respectively.
%
Therefore the Coulomb energy contribution to the total energy is proportional to
the squared total number of particles and local moment,
$E_{U} = \frac{1}{2} \bar{U} \langle N^2  \rangle - \frac{1}{4}I \langle m_z^2 \rangle$.
%
Since $\langle N^2 \rangle$ only weakly depends on the $c/a$ distortion, the main
dependence of the total energy of the paramagnetic phase on $c/a$ is due to the
magnetic correlation energy.
\end{thebibliography}
\end{document}